\advance\hoffset -0.5cm
\advance\voffset -1.0cm
\documentstyle[12pt]{article} 

\newlength{\height}
\divide \height by 3

\begin{document} 
\newcommand{\be}{\begin{equation}}
\newcommand{\ee}{\end{equation}}
\newcommand{\bea}{\begin{eqnarray}}
\newcommand{\eea}{\end{eqnarray}}
\newcommand{\cl}{{\cal L}}
\newcommand{\cd}{\not\!\! D}
\newcommand{\req}[1]{(\ref{#1})}
\newcommand{\qt}{QED$_3$}
\newcommand{\ink}{\int {d^3k\over(2\pi)^3}}
\newcommand{\al}{{\alpha\over16}}
\newcommand{\pe}{\Pi_{\rm even}}
\newcommand{\po}{\Pi_{\rm odd}}
\newcommand{\wn}{{\widetilde N}}
\newcommand{\wnc}{{\widetilde N_c}}

\def\spur#1{\mathord{\not\mathrel{#1}}}
\baselineskip=\height 
\begin{titlepage}
\begin{center}
\makebox[\textwidth][r]{SNUTP 93-48}
\vskip 0.35in
{{\Large \bf DYNAMICAL EFFECTS OF CHERN-SIMONS
TERM\footnote[1]{To be published in the proceedings of the Mount
Sorak Winter School, Korea}}}
\end{center}
\begin{center}
\par \vskip .1in \noindent Deog Ki Hong\footnote[2]{Supported in part
by the Korean Science and Engineering Foundation
through the SRC program of SNU-CTP.}
\end{center}

\begin{center}
Department of Physics, Pusan National University\\ Pusan
609-735, Korea\\


\end{center}

\begin{abstract}
We study how the Chern-Simons term effects the dynamically
generated fermion mass in $(2+1)D$ Quantum Electrodynamics in
the framework of large $N$ expansion. We find that when the
Chern-Simons term is present half of the fermions get mass $M+m$
and half get $M-m$.  The parity-preserving mass $m$ is generated
only when $N < \wnc$.  Both the critical number, $\wnc$, of
fermion flavor and the magnitude, $m$, reduce when the effect of
the Chern-Simons term dominates.

\end{abstract}
\end{titlepage}

\section{Motivation}

One of the most drastic effects of the Chern-Simons term, ${\cal
L}_{CS}=\kappa \epsilon_{\mu\nu\lambda} A^{\mu}
\partial^{\nu}A^{\lambda}$, in field theories,
is the generation of fractional spin.  The Chern-Simons term
induces fractional spin to a particle coupled to the
Chern-Simons gauge field, $A_{\mu}$ \cite{wilczek}: the spin is
given as
\be
s= {e^2 \over 4\pi \kappa}.
\label{spin}
\ee
Due to this effect, the Chern-Simons theory is not only
interesting, field theoretically, but also has some applications
in the condensed matter systems like the fractional or integer
quantum Hall system \cite{qhe}.  The Chern-Simons term is
topological in a sense that it does not involve a metric and
thus does not contribute to energy-momentum tensor of the
system. But, it modifies the equations of motion and breaks
parity and Time-reversal symmetry, which must have dynamical
significances.  One nice example for this dynamical effect is
the existence of a stable vortex solution found recently
\cite{yoonbai} in a system described by a
Lagrangian density,
\be
{\cal L}={1\over
2}\left|D_{\mu}\phi\right|^2-V(\left|\phi\right|)+{\cl}_{CS},
\label{vortex}
\ee
where $D_{\mu}=\partial_{\mu}-ieA_{\mu}$ and
$V(\left|\phi\right|)$ is a $\phi^6$ potential. In
(2+1)-dimensions, the kinetic energy of the static soliton is
scale-invariant, while the potential energy is not. Therefore,
without a gauge field or other balancing force, the static
soliton is unstable against collapsing to the center of the
soliton \cite{jkps}.  Namely, it is energetically favorable for
the soliton to collapse to $\phi=v$, where $v$ is the minimum of
the potential. One would think that adding the Chern-Simons term does
not do any good in stabilizing the soliton, since it does not
contribute to the energy of the soliton. But, this is not true,
since not only the energy of the soliton has now a term
depending on the gauge field,
${1\over2}A_{\mu}^2\left|\phi\right|^2$, but also the
Chern-Simons term modifies the equation of motion, and thus a
stable vortex solution is possible.

In this talk, I would like to present another effect of the
Chern-Simons term \cite{hong1}, namely the dynamical effect of
the parity-noninvariance of the Chern-Simons term.

\section{Parity and Mass in $(2+1)$-dimensions}

Consider a $(2+1)D$ QED, described by a Lagrangian density
\be
\cl=\bar \psi i \cd \psi -{1\over4} F_{\mu\nu}^2+\cl_{\rm mass}
\label{qed}
\ee
where $\psi$ is a two-component spinor and the gamma matrices
are chosen as $\gamma^0=\sigma^3, \: \gamma^1=i\sigma^1, \:
\gamma^2=i\sigma^2$. The
mass terms for the fermion and the photon are
\be
\cl_{\rm mass}=-m\bar\psi\psi+
\kappa\epsilon_{\mu\nu\lambda}A^{\mu}\partial^{\nu}A^{\lambda}
\label{mass}
\ee

In (2+1)-dimensions, parity is defined to be a coordinate
transformation, $P:\; x=(x,y,t)\mapsto x^{\prime}=(-x,y,t)$,
under which the fields transform as following:
\bea
A^{0,2}(x) & \mapsto & {A^{\prime}}^{0,2}(x^{\prime})=A^{0,2}(x)
\nonumber \\
A^1(x) & \mapsto & {A^{\prime}}^1(x^{\prime})= -A^1(x) \\
\psi(x) & \mapsto & \psi^{\prime}(x^{\prime})=e^{i\delta}\sigma^1\psi(x)
\label{parity}
\eea

One can therefore easily see that the both mass terms are odd
under parity (and also under time-reversal). If either of the
mass terms is absent at tree level, it will be generated
radiatively, since the parity, which forbids the mass term, is
broken by the other mass term explicitly.  For example, when the
(topological) mass term for the gauge field is absent, the
fermion mass term will generate it radiatively with a coefficient
$\kappa={e^2\over 8\pi} {m \over |m|}$ \cite{redlich}.
Similarly, when the fermion mass term is absent, the
Chern-Simons term will generate it at one-loop; one needs a
counter-term to remove the divergence in the fermion mass,
$\delta m= -{6\over\pi}{e^2\over\kappa}|M|$, where $M$ is the
Pauli-Villas regulator \cite{hong2}.

When the number of the fermion flavors is even, the system has
another obvious discrete symmetry, $Z_2$, which interchanges
half of the fermions with another half; for $i=1,\cdots,
{N\over2}$, $Z_2$ mixes the fermions fields as
\bea
\psi_i(x) & \mapsto & \psi_{{N\over2}+i}(x) \nonumber \\
\psi_{{N\over2}+i} & \mapsto & \psi_i(x)
\label{z2}
\eea
If we define a new parity, $P_4\equiv PZ_2$, combining the old
one with $Z_2$, then the fermions can have
``parity($P_4$)-invariant" mass, $m_i\bar\psi_i\psi_i$, with
\be
 m_i=\left\{ \begin{array}{ll} m, & \mbox{if $1\le i \le {N\over2}$} \\
                             -m, & \mbox{if ${N\over2}+1\le i \le N$}
            \end{array}
    \right.
\ee
With this form of fermion mass, the Chern-Simons term will not
be generated radiatively. We call this ``parity($P_4$)-even
mass".  On the other hand, this $P_4$-invariant fermion mass can
be generated dynamically due to a non-perturbative effect,
though $\cl_{\rm mass}$ is not present in the Lagrangian; namely
$P$ is spontaneoulsy broken, while $P_4$ is not.  Appelquist et.
al. \cite{appelquist} showed, using $1/N$-expansion, that, if
$1/N > 1/N_c$ with $N_c=32/\pi^2$, the fermions condensate and
thus the parity-even fermion mass is generated dynamically and the
mass is given as
\be
m_{\rm even}= \alpha e^{-{\pi^2\over 16}/{1\over N}}
\ee

When the Chern-Simons term is present, this parity-even mass
will be affected. As described below, due to the Chern-Simons
term, $1/N_c$ increases (one needs a stronger interation to form
a fermion condensate) and the magnitude of the parity-even
fermion mass decreases.

\section{Gap Equation}

First, we will examine the pattern of the spontaneous breaking
of parity in the $(2+1)D$ QED with $N$ complex two-component
spinors, and then elaborate on the dynamical mass generation for
fermions.  An order parameter for the spontaneous breaking of
parity is the vacuum condensate of the fermion bilinear,
$\left<\bar\psi\psi(x)\right>$, which can be determined once one
finds the asymptotic behavior of the fermion propagator
\cite{georgi}.
We use the $1/N$-expansion technique, since it is a
non-perturbative technique and also the IR-divergence of
$(2+1)D$ QED softens in the large $N$ limit \cite{appelquist1}.

At the leading order in the $1/N$ expansion, the Dyson-Schwinger
gap equation is, in Euclidean notation,
\be
-(Z(p)-1)\not\!p+\Sigma(p)={\alpha\over N}
\int {d^3k\over(2\pi)^3}D_{\mu\nu}(p-k)\gamma_{\nu}
{Z(k)\not\!k-\Sigma(k)\over Z^2(k)k^2+\Sigma^2(k)}\gamma_{\mu},
\label{gap}
\ee
where $D_{\mu\nu}^{-1}=\Delta_{\mu\nu}^{-1}-\Pi_{\mu\nu}$ and
$\Delta_{\mu\nu}$ is the free Landau gauge propagator and
$\alpha\equiv e^2N$ is kept fixed while $N\to\infty$.  $\Sigma$
is the fermion self energy.  The vacuum polarization tensor is
given as
\be
\Pi_{\mu\nu}   =  \Pi_{\rm even}(p)
\left(\delta_{\mu\nu}-{p_{\mu}p_{\nu}\over p^2}\right)
+\Pi_{\rm odd}(p)\epsilon_{\mu\nu\lambda}p_{\lambda}.
\ee
{}From the equation (\ref{gap}), taking trace over the gamma
matrix, we get
\be
\Sigma(p)={\alpha\over N} \ink {2\Pi_1(p-k)\over (p-k)^2}
{\Sigma(k)\over k^2+\Sigma^2(k)}+ {\alpha\over N} \ink
{(p-k)\cdot k\over \left|p-k\right|^3} {\Pi_2(p-k)\over
k^2+\Sigma^2(k)}
\label{mas}
\ee
where $\Pi_1$ and $\Pi_2$ contain the quantum corrections to the
parity-even part and the parity-odd part of the photon
propagator;
\be
D_{\mu\nu}(p)={\delta_{\mu\nu}-p_{\mu}p_{\nu}/p^2\over p^2}
\Pi_1(p)+
{\epsilon_{\mu\nu\lambda}p^{\lambda}\over
\left|p\right|^3}\Pi_2(p),
\ee
with
\bea
\Pi_1(p) & = & {1-\pe(p)/p^2\over
(\pe(p)/p^2)^2+(\kappa-\po(p))^2/p^2} \\ \Pi_2(p) & = &
-{(\kappa-\po(p))/|p|\over (\pe(p)/p^2)^2+(\kappa-\po(p))^2/p^2}
\eea
In the large-$N$ approximation the dynamically generated mass
will be at most of order of $1/N$, compared to the scale of the
theory, $\Lambda$.  ($\Lambda$ is of same order as $\alpha$ or
$\kappa$.) In the region, $\Sigma(p)\ll p$, the vacuum
polarization tensor takes a simple form;
\bea
\Pi_{\rm even}(p) & = & -{\alpha\over 16}\left|p\right| \\
\Pi_{\rm odd}(p)  & = & {1\over N}\sum_{i=1}^NM_i
{\alpha\over 4 \left|p\right|}
\eea
where $M_i\simeq\Sigma_i(0)$, the mass of the $i$-th fermion.
(In general $\Sigma(p)$ will depend on the flavor but we will
suppress the index $i$ for simplicity.)

To find the physical mass, we have to solve
\be
p^2+\Sigma^2(p)=0 \;\;{\rm at}\;\; p^2=-m_{\rm phy}^2
\ee
But, since $\Sigma(p)$ is quite small, compared to the scale of
the theory ($\alpha$ or $\kappa$), we may take
\be
m_{\rm phy}\simeq \Sigma(0).
\ee
Then, the gap equation (\ref{gap}) becomes
\be
\Sigma_i(0)={\alpha\over N}\int{d^3k\over(2\pi)^3}
\left({2\Pi_1(k)M_i\over k^2(k^2+M_i^2)}-
{\Pi_2(k)\over |k|(k^2+M_i^2)}\right)
\label{mass1}
\ee
Note that the first term depends on the flavor while the second
term does not. The parity ($P_4$) is maximally broken when the
second term is dominant, which happens precisely when the
Chern-Simons term is dominant.  On ther other hand, if
Chern-Simons term is not present, the mass will be generated in
a parity-invariant way. Namely, half the fermions get positive
mass $m$ and the other half negative mass $-m$.  Therefore, when
both of the Chern-Simons term and the Maxwell term are present,
it is reasonable to assume the pattern of the fermion mass as
\be
M_i=M+m_i,
\ee
with
\be
m_i=\left\{ \begin{array}{cl} m, &\mbox {$i=1,\ldots, N-L$},\\
-m, &\mbox {$i=N-L+1,\ldots,N$} \end{array} \right.
\ee
By plugging this Ansatz into Eq. (\ref{mass1}), we obtain
\be
M = {\alpha\over 2\pi^2N}\int^{\Lambda}_{m}dk {k^2\over
(k^2+M^2)}{\kappa\over \left(\al\right)^2+\kappa^2} \simeq
{\Lambda\over 2\pi^2N}{\alpha\kappa\over(\al)^2+\kappa^2}
\label{main2}
\ee
and
\be
m_i = {1\over \pi^2N} {\left(\al\right)^2\over
\left(\al\right)^2+\kappa^2}
\int_m^{\Lambda}dk \left({16m_i\over k}-{64\over k}\theta m
\right),
\label{main1}
\ee
where $\theta=1-2L/N$.  For Eq.(\ref{main1}) to have a
consistent solution, $\theta=0+O(1/N)$, which yields, upon
integration,
\be
m = \Lambda\exp(-4N/\wnc)
\ee
where $\wnc=N_c/(1+({16\kappa\over\alpha})^2)$.  The value for
the parity-violating mass $M$ is a perturbative one in the $1/N$
expansion, while the parity-preserving mass is nonperturbative.
The effect of the Chern-Simons term is now clear.  It induces a
parity-violating mass perturbatively and it decreases in a
nonperturbative way the magnitude of the parity-preserving mass
$m$.  Since $\theta=0$, half the fermions get (positive) mass
$M+m$ and the other half $M-m$. The pattern of the
flavor-symmetry breaking is same whether the Chern-Simons term
is present or not.

\section{Conclusion}

In conclusion, we find that, when the Chern-Simons term is
present, the parity tends to be maximally broken; the magnitude
of the parity-even (four-componet) mass for the fermions gets
smaller, and the critical number for the generation of the
parity-even mass decreases. But, in the large-$N$ limit, the
pattern of the flavor-symmetry breaking does not depend on the
Chern-Simons term.


\pagebreak

\end{document}